\documentclass[preprint,12pt]{elsarticle}

\usepackage{lineno}
\usepackage{multirow}
\journal{Nucl. Instrum. Methods A}

\begin{document}

\begin{frontmatter}



\title{Noise discrimination method based on charge distribution of CMOS detectors for soft X-ray}


\author[a,b]{Xinchao Fang}
\author[c]{Jirong Cang}
\author[c]{Qiong Wu}
\author[b,c]{Hua Feng\corref{cor}}
\ead{hfeng@tsinghua.edu.cn}
\author[a,b]{Ming Zeng\corref{cor}}
\ead{zengming@tsinghua.edu.cn}
\cortext[cor]{Corresponding author}

\affiliation[a]{organization={Key Laboratory of Particle and Radiation Imaging (Tsinghua University)},
            addressline={Ministry of Education},
            city={Beijing},
            postcode={100084},
            country={China}}

\affiliation[b]{organization={Department of Engineering Physics},
            addressline={Tsinghua University},
            city={Beijing},
            postcode={100084},
            country={China}}
            
\affiliation[c]{organization={Department of Astronomy},
            addressline={Tsinghua University},
            city={Beijing},
            postcode={100084},
            country={China}}

\begin{abstract}
Complementary metal-oxide semiconductor (CMOS) sensors have been widely used as soft X-ray detectors in several fields owing to their recent developments and unique advantages. The parameters of CMOS detectors have been extensively studied and evaluated. However, the key parameter signal-to-noise ratio in certain fields has not been sufficiently studied. In this study, we analysed the charge distribution of the CMOS detector GSENSE2020BSI and proposed a two-dimensional segmentation method to discriminate signals according to the charge distribution. The effect of the two-dimensional segmentation method on the GSENSE2020BSI dectector was qualitatively evaluated. The optimal feature parameters used in the two-dimensional segmentation method was studied for G2020BSI. However, the two-dimensional segmentation method is insensitive to feature parameters.
\end{abstract}

\begin{graphicalabstract}
\includegraphics[width=8cm]{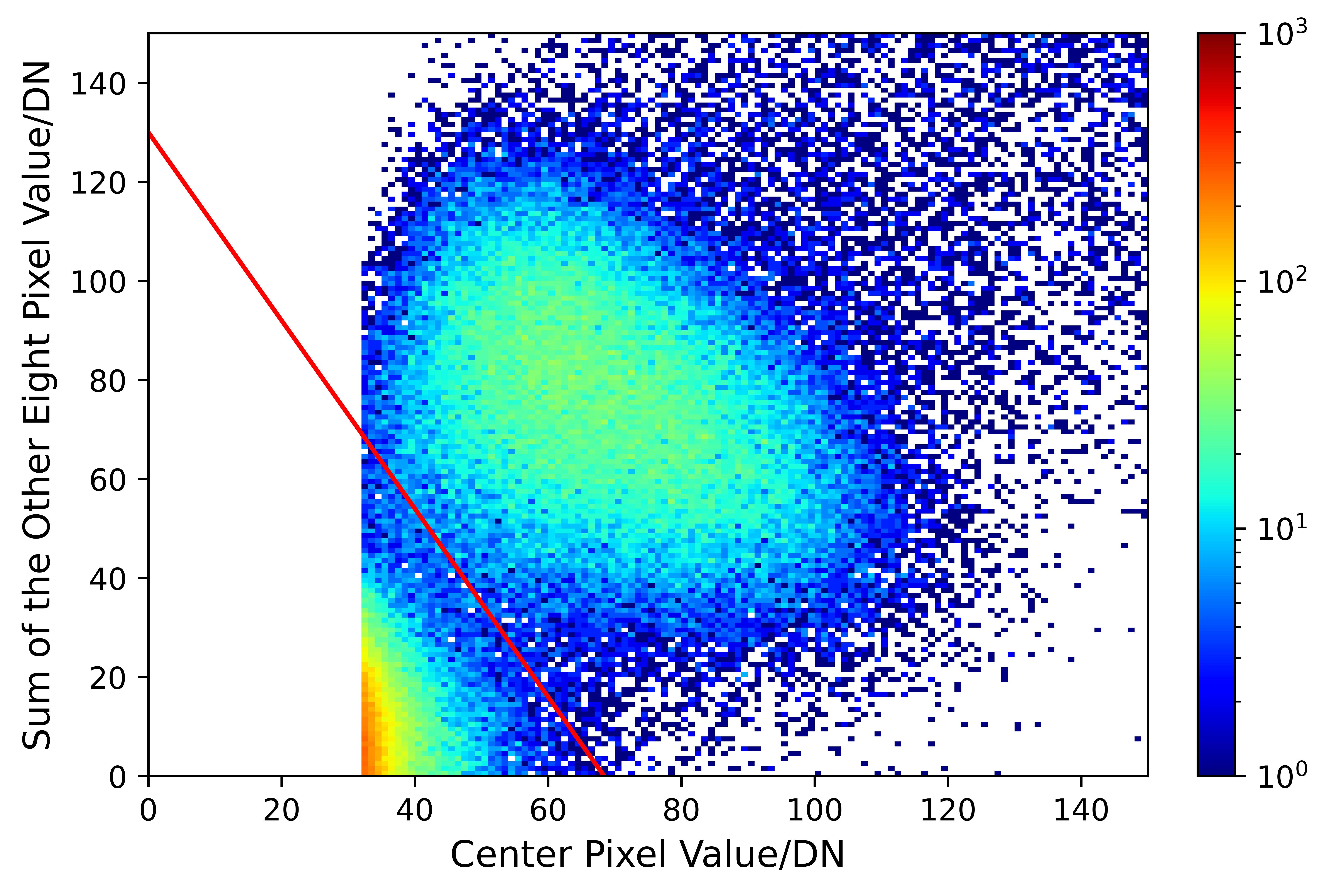}
\par{the mixed dataset include all single-pixel and multi-pixel events.\par}
\includegraphics[width=8cm]{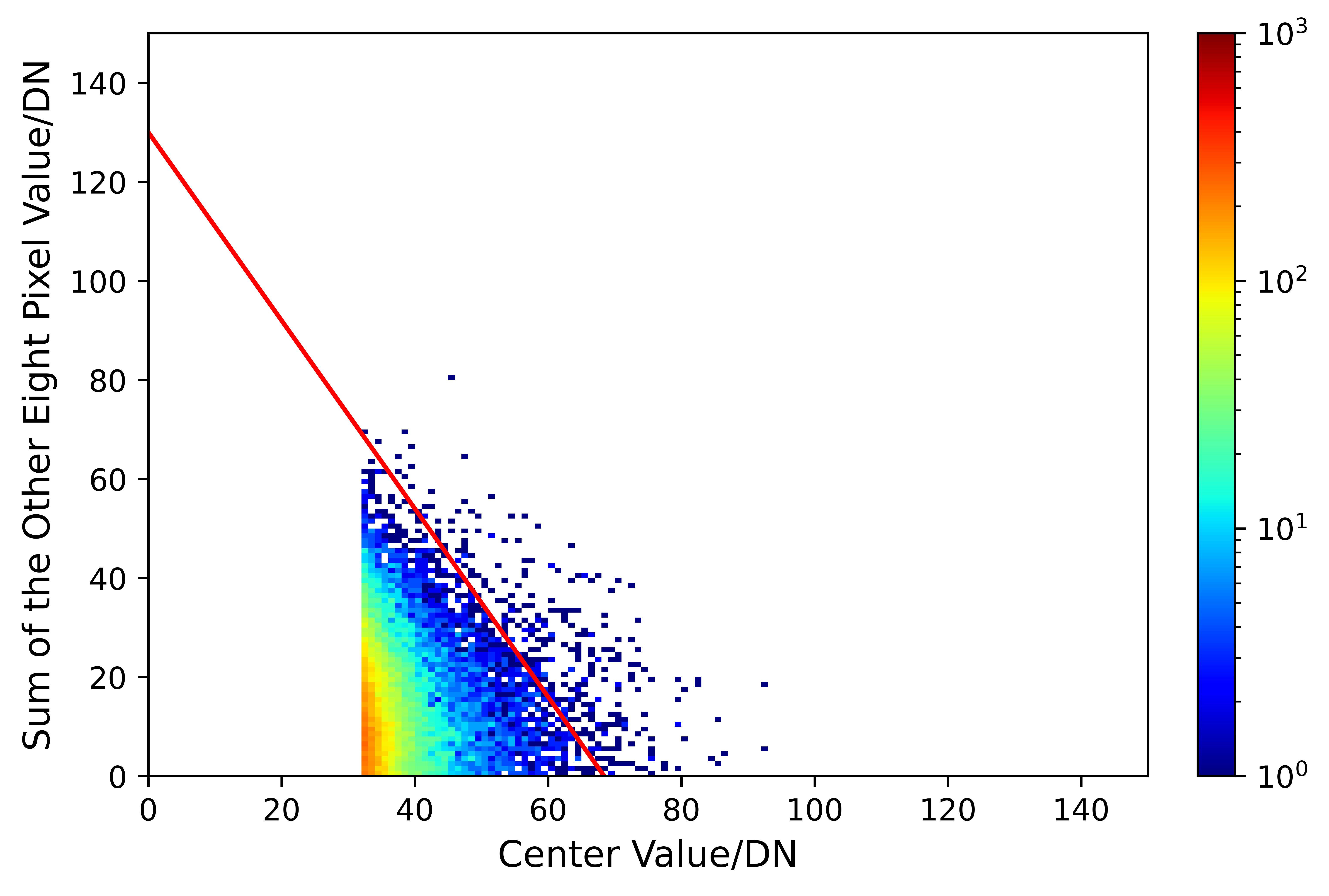}
\par{the noise dataset include all single-pixel and multi-pixel events.\par}
\includegraphics[width=8cm]{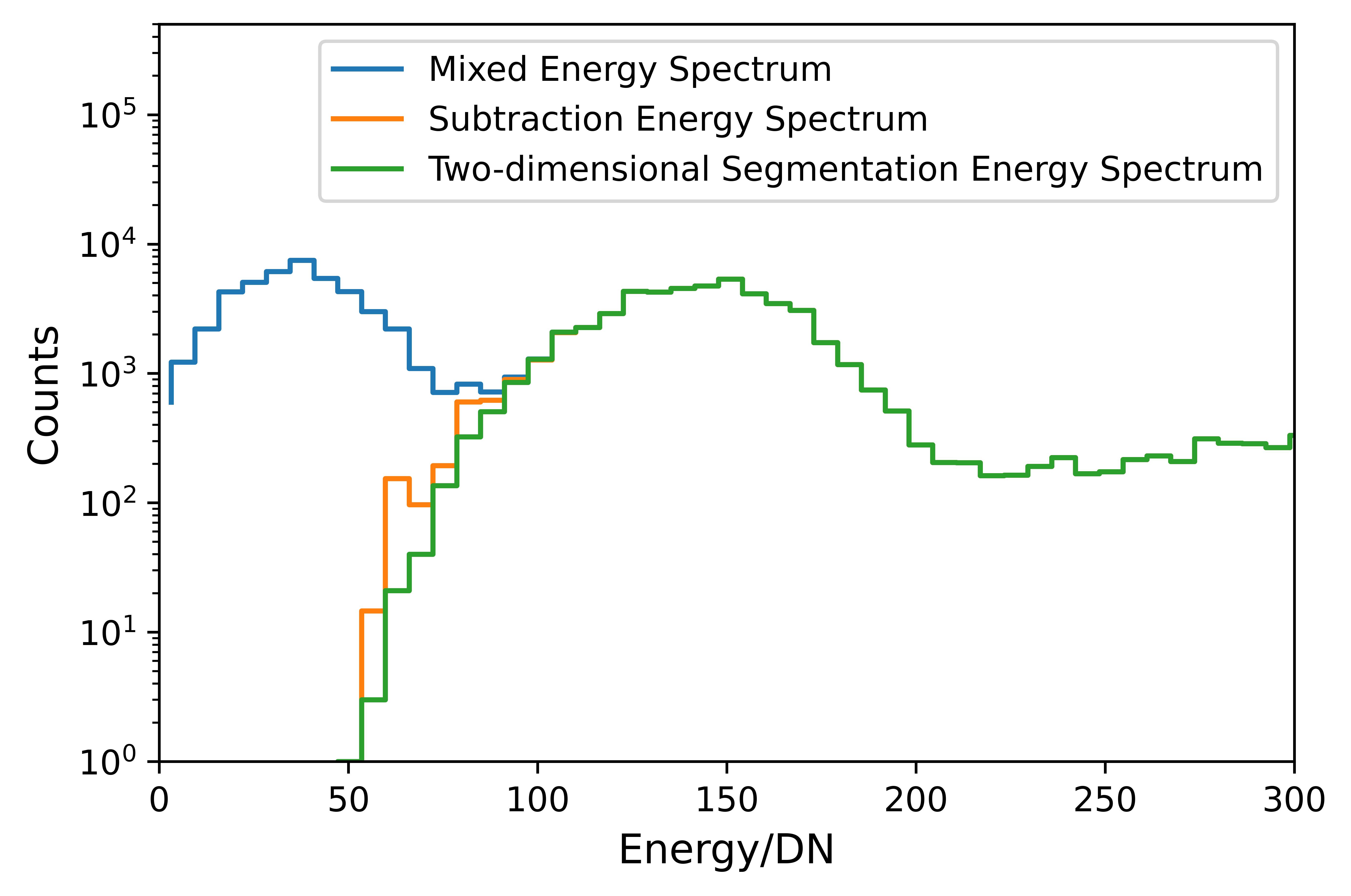}
\par{Energy spectrum plotted on the mixed dataset processed with two-dimensional segmentation method.\par}
\end{graphicalabstract}

\begin{highlights}
\item a new noise-discrimination methods based on the charge distribution of CMOS detector
\item this method is insensitive to feature parameters, and good results can be achieved when using empirically selected feature parameters.
\item CMOS detector GSENSE2020BSI with energy lower threshold lower the K$_{\alpha}$ line of Carbon (277 eV) at room temperature without multi-pixel events loss
\end{highlights}

\begin{keyword}
CMOS detector \sep Soft X-ray detector \sep Charge distribution \sep Noise discrimination



\end{keyword}

\end{frontmatter}


\section{Introduction}
    Owing to the rapid development of backside illumination (BSI) complementary metal-oxide semiconductor (CMOS) sensors in recent years \cite{Harada_2019,Harada_2020}, CMOS detectors have several excellent properties such as small pixel size, high quantum efficiency, low readout noise, wide work temperature, high time resolution,and low cost. This makes CMOS detectors possible alternatives to charge-coupled device (CCD) detectors for soft X-ray detection in several fields, and scientific results have already been published \cite{Cooper2020,Desjardins2020,Ishikawa2018191,Julian2020}. Therefore, it is important to study and optimize the performance of CMOS detectors.\par
    Numerous studies have been concluded to optimise the energy resolution of some CMOS detectors, and progress has been achieved. The influence of energy resolution, such as peak shift \cite{Narukage2020}, different gains between pixels \cite{Wu2023a}, and image lag \cite{Wu2023}, have been considered and extensively studied. However, in addition to energy resolution, the signal-to-noise ratio (SNR) significantly contributes to the sensitivity of the detector. SNR refers to the ratio of the number of signal events to the number of detector noise events. In astronomy, the higher the SNR of a detector, the lower the minimum detectable polarisation, which indicates that the detector is more capable of detecting polarisation \cite{Weisskopf2010}. Therefore, it is worth more attention than before to the SNR. \par
    In this study, the charge distribution of CMOS detectors was studied to obtain a better SNR, and a new noise-discrimination methods based on the charge distribution was proposed. In particular, in this study, we concluded the charge distribution and noise-discrimination method of a GSENSE2020BSI (G2020BSI) detector because the GSENSE series of back-illuminated CMOS detectors has been widely employed in several fields \cite{Ogino2021,Chen2022,Mille2022} owing to their excellent energy, spatial, and temporal resolutions. \par
    
\section{The G2020BSI Detector and its Charge Distribution Model}
    Scientific CMOSs based on a standard four-transistor (4T) pixel architecture \cite{Zhang2022,Heymes2022,Ma2015} are widely used in soft X-ray astronomy. The detector used in our study, G2020BSI is a typecal one with sufficiently good quantum efficiency ( $>$ 90\% in the range of 80-1000 eV). The G2020BSI sensor contains a 2048$\times$2048 array of 6.5 square $\mu$m back-illuminated pixels and is capable of operating at a maximum frame rate of 74 fps. According to the G2020BSI manual, the temporal dark noise is 1.6 $e^{-}$ at 25 $^{\circ}$C, which is 2.5 $e^{-}$ in our measurement. In our experiments, we used an evaluation board of the G2020BSI provided by Gpixel. To evaluate the basic performance of G2020BSI, we built a vacuum system as shown in Figure 1. A carbon-target X-ray light tube, Mo-target X-ray light tube and an $^{55}$Fe radiation source were used as the soft X-ray sources. For these radiation sources, we obtained 400, 100, and 6000 images at room temperature, 10$^{-5}$ Pa with an exposure time of 22.4 ms, respectively. The numbers of images of different sources were varied to ensure that the number of major energy peak events in the spectrum was comparable for different source intensities and that the errors owing to statistical properties were reasonable. In addition, we measured 1000 images in a dark field at room temperature, 10$^{-5}$ Pa with an exposure time of 22.4 ms.\par
    The data processing method has been studied in the relevant literature\cite{Narukage2020,Wang2019,Desjardins2020}. We implemented our code, which can be briefly described as follows. First, we filtered the hot pixels (i.e. pixels with high noise levels). We obtained 1000 noise images by subtracting the average dark field image from 1000 dark field images. The standard deviation and maximum noise for each pixel were calculated statistically, and the two-dimensional distribution of the standard deviation of noise versus the maximum noise for each pixel of the sensor is shown in Figure 2. Some pixels have small standard deviation of noise, but a large maximum noise. A typical noise distribution of such pixels, which may have been caused by the internal structure of the sensor, is shown in Figure 3. We do not count the events incident on pixels with noise maxima greater than 31 digital numbers (DN), which we consider to be hot pixels. The spatial distribution of hot pixels with a hot pixel threshold of 31 DN is shown in Figure 4, which accounts for 3.4\% of the total 2048$\times$2048 pixels.\par
    Each X-ray images was subtracted from the corresponding baseline image. The single-photon measurements used throughout the measurement means that the majority of pixels in each image did not receive photon events. Moreover, the temperature changes should be small over a short period. Therefore, for each X-ray map, the median image of the first 10 images was considered as the baseline image of the processed image, which can reduce the effects of temperature and image lag. For instance, the median image of the (n-10)$_{th}$ to the (n-1)$_{th}$ image is the baseline image of the n$_{th}$ image. \par
    We considered that the energy of the photon event was mainly deposited on the incident pixels and that the energy did not spread beyond the 3$\times$3 pixel region. The sum of the pixel energies in the 3$\times$3 region was selected as the energy of the event. The threshold values were set to T$_{Carbon}$= 31 DN, T$_{Mo}$ = 94 DN, and T$_{Fe}$ = 94 DN. For all over-threshold events, events with the energy of the center pixel greater than 60\% of the event value were selected as single-pixel events. In other words, the selected events had less energy loss and smaller energy-spectrum aberrations occurred in the spectrum of these selected events. As shown in Figure 5, the characteristic X-rays corresponding to the main energy peaks including K$_{\alpha}$ line of Carbon (277 eV), K$_{\alpha}$ line of Mo (2293 eV), and K$_{\alpha}$ line of Mn (5899 eV) could be clearly observed in the energy spectrum. While the characteristic X-rays corresponding to the other energy peaks including K$_{\alpha}$ line of O, K$_{\alpha}$ line of Si (1740 eV), K$_{\beta}$ line of Mn (6490 eV), and Si escape peak of Mn (4159 eV) could be observed despite the insufficient statistics. The energy calibration curve plotted according to the positions of the energy peaks is shown in Figure 6. The results obtained by straight-line fitting are shown in Equation (1). \par
    \begin{figure}[hbt]
        \centering
        \includegraphics[width=8cm]{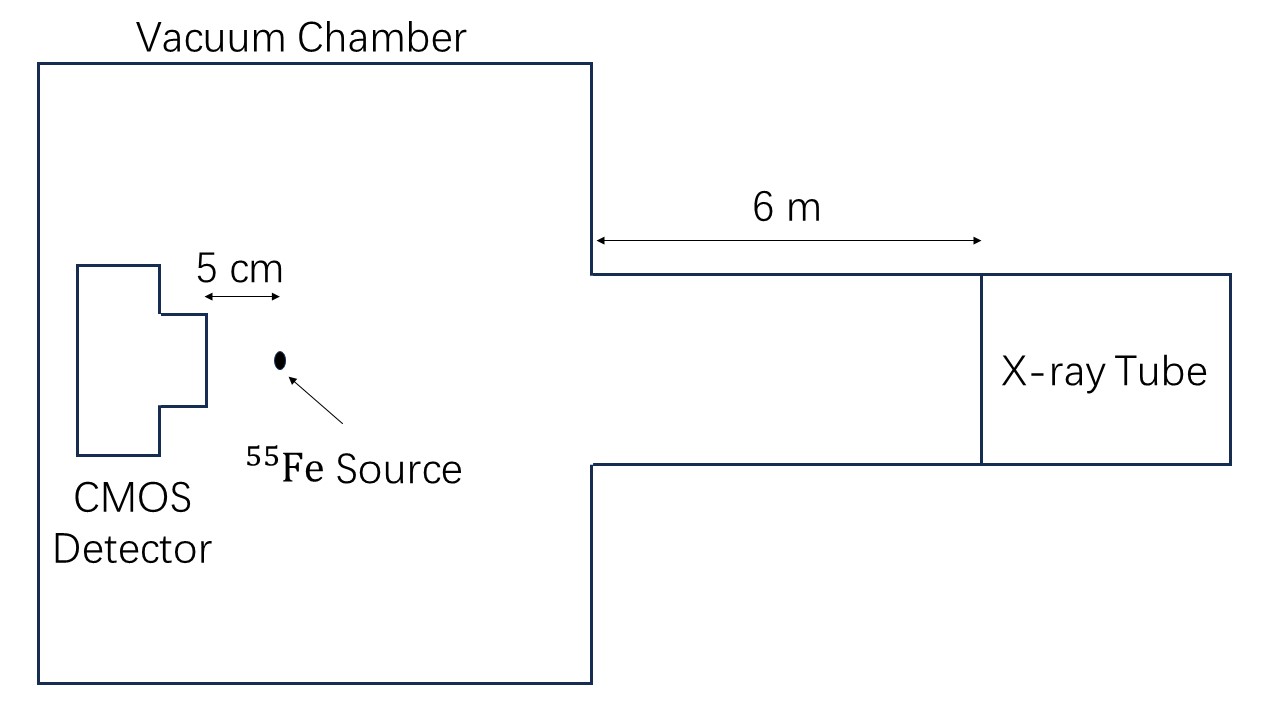}
        \caption{The vacuum system used for characteristic X-ray experiment in G2020BSI detector}\label{Figure 1}
    \end{figure} 
    \begin{figure}[hbt]
        \centering
        \includegraphics[width=8cm]{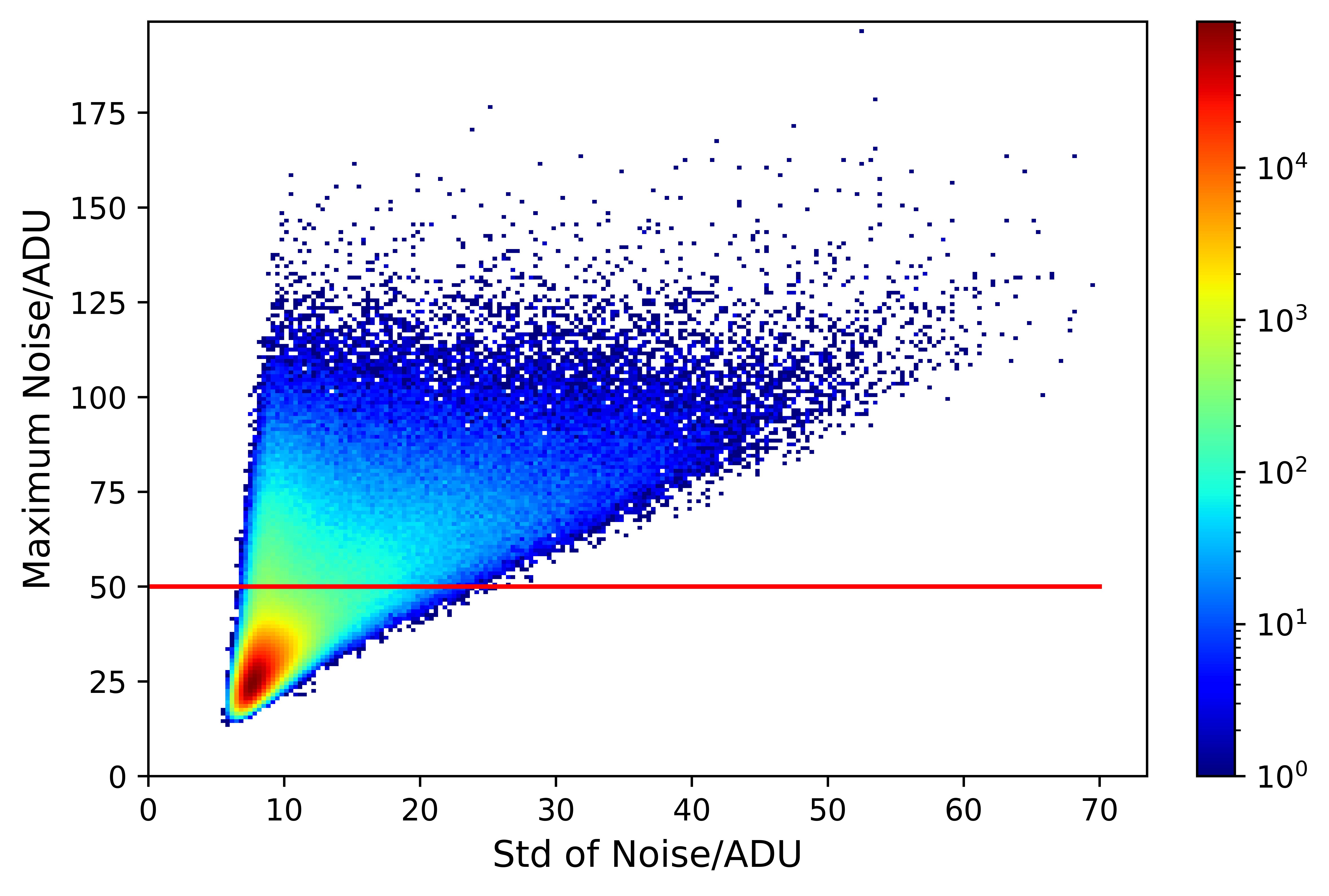}
        \caption{The two-dimensional distribution of noise std vs. noise maximum of the same pixel. The red line is the hot pixel threshold.}\label{Figure 2}
    \end{figure} 
    \begin{figure}[hbt]
        \centering
        \includegraphics[width=8cm]{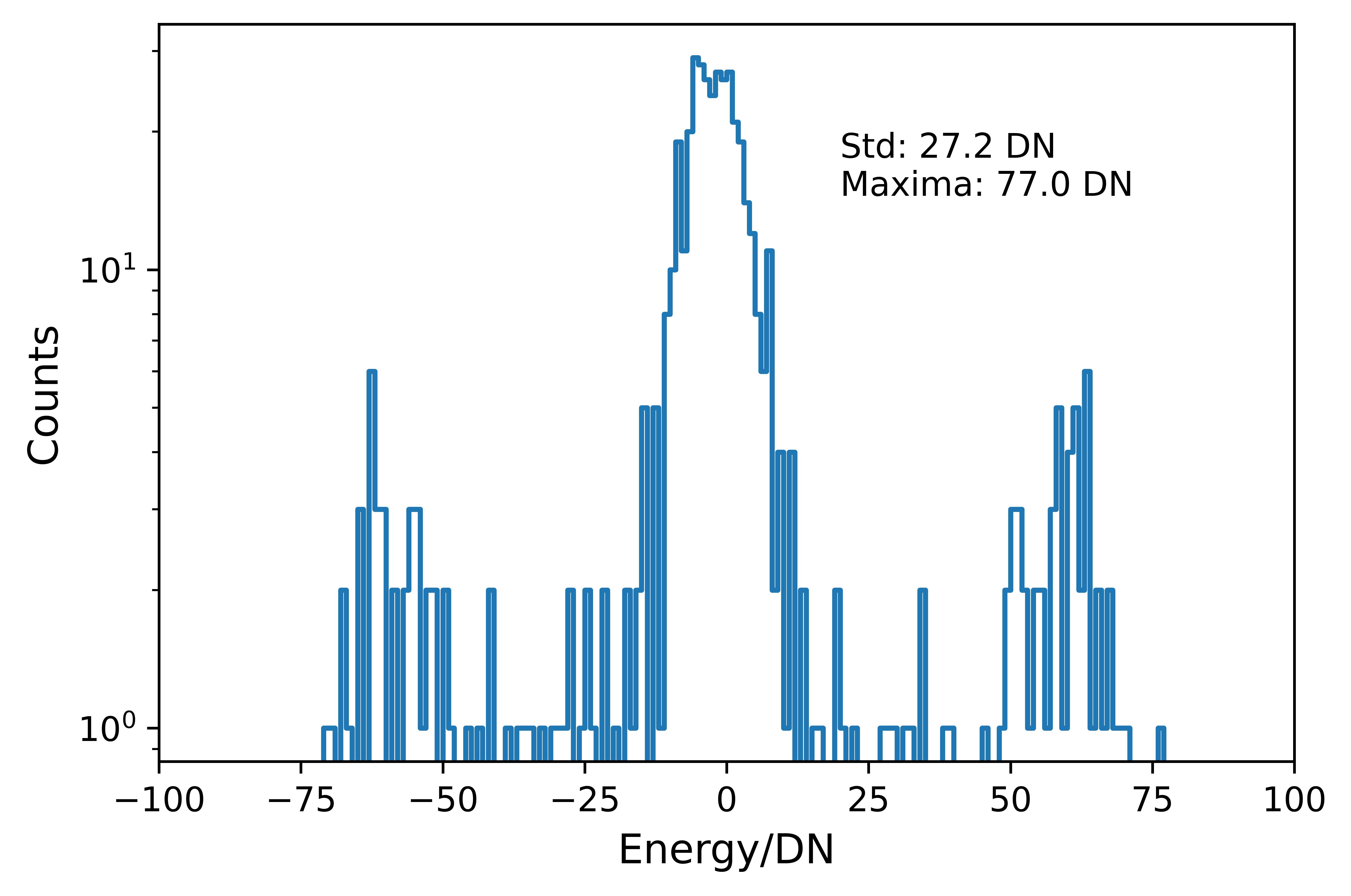}
        \caption{Dark noise distribution of pixels in row 1087 and column 120 at room temperature.}\label{Figure 3}
    \end{figure} 
    \begin{figure}[hbt]
        \centering
        \includegraphics[width=8cm]{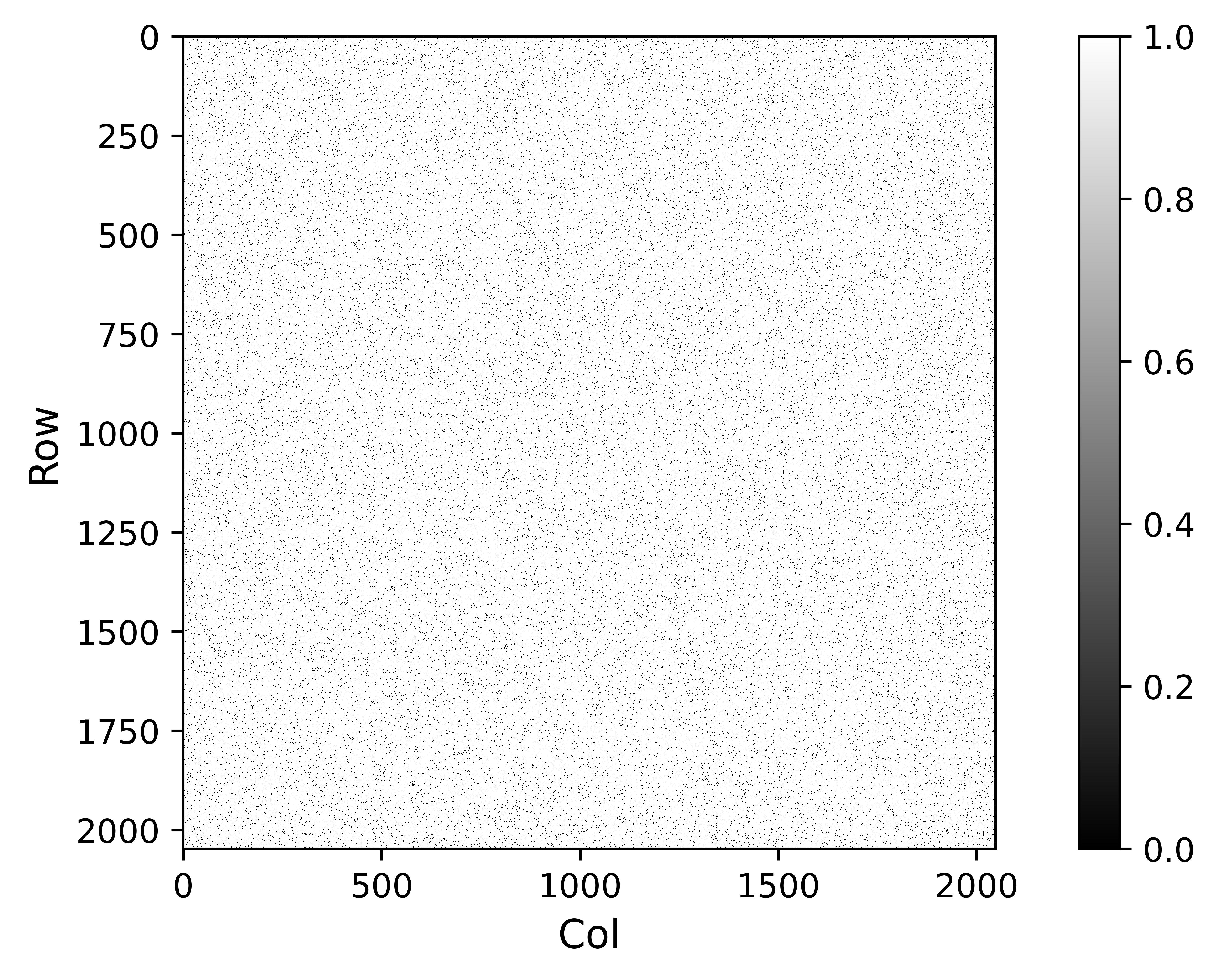}
        \caption{The spatial distribution of hot pixels is essentially uniform, and hence, the idea of finding a large enough chip area completely free of hot pixels for the experiment is discarded.}\label{Figure 4}
    \end{figure}

    \begin{equation}\label{Equation (1)}{}
    \centering
    E\ =\ 1.57N + 68.17
    \end{equation}
    where E and N are the event energy in eV and DN, respectively. The slope of the linear fit was 1.57 eV/DN, considering that the average electron-hole pair ionization energy in the Si detector was 3.65 eV/$e^-$, and the conversion factor was 0.430 $e^-$/DN. The accuracy of the energy spectrum was verified using silicon drift detectors.\par 
    \begin{figure}[hbt]
        \centering
        \includegraphics[width=8cm]{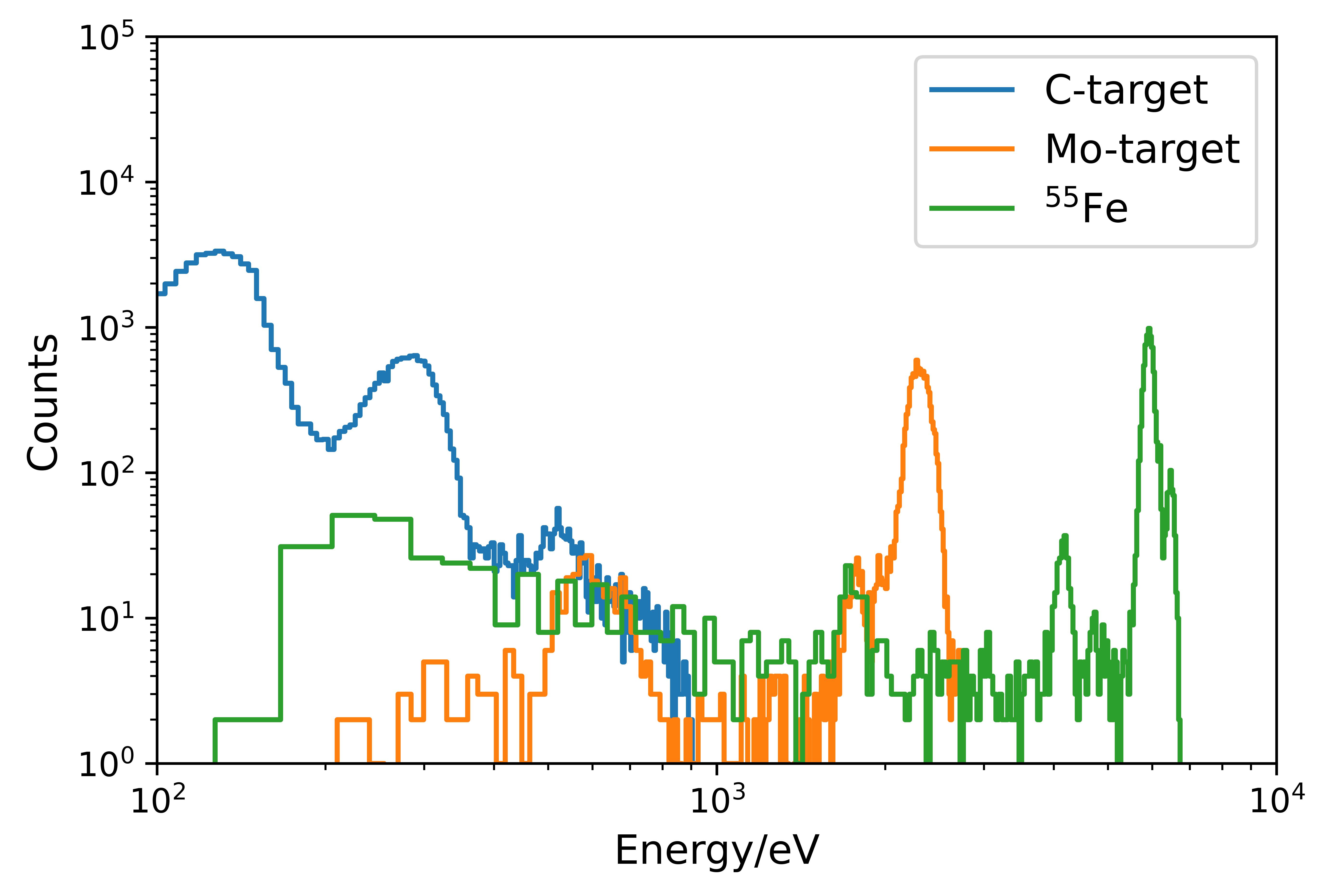}
        \caption{The spectrum with different sources. The blue line indicates the spectrum for carbon target images. The orange line indicates the spectrum for Mo target images. The green line indicates the spectrum for $^{55}$Fe target images.}\label{Figure 5}
    \end{figure}
    \begin{figure}[hbt]
        \centering
        \includegraphics[width=8cm]{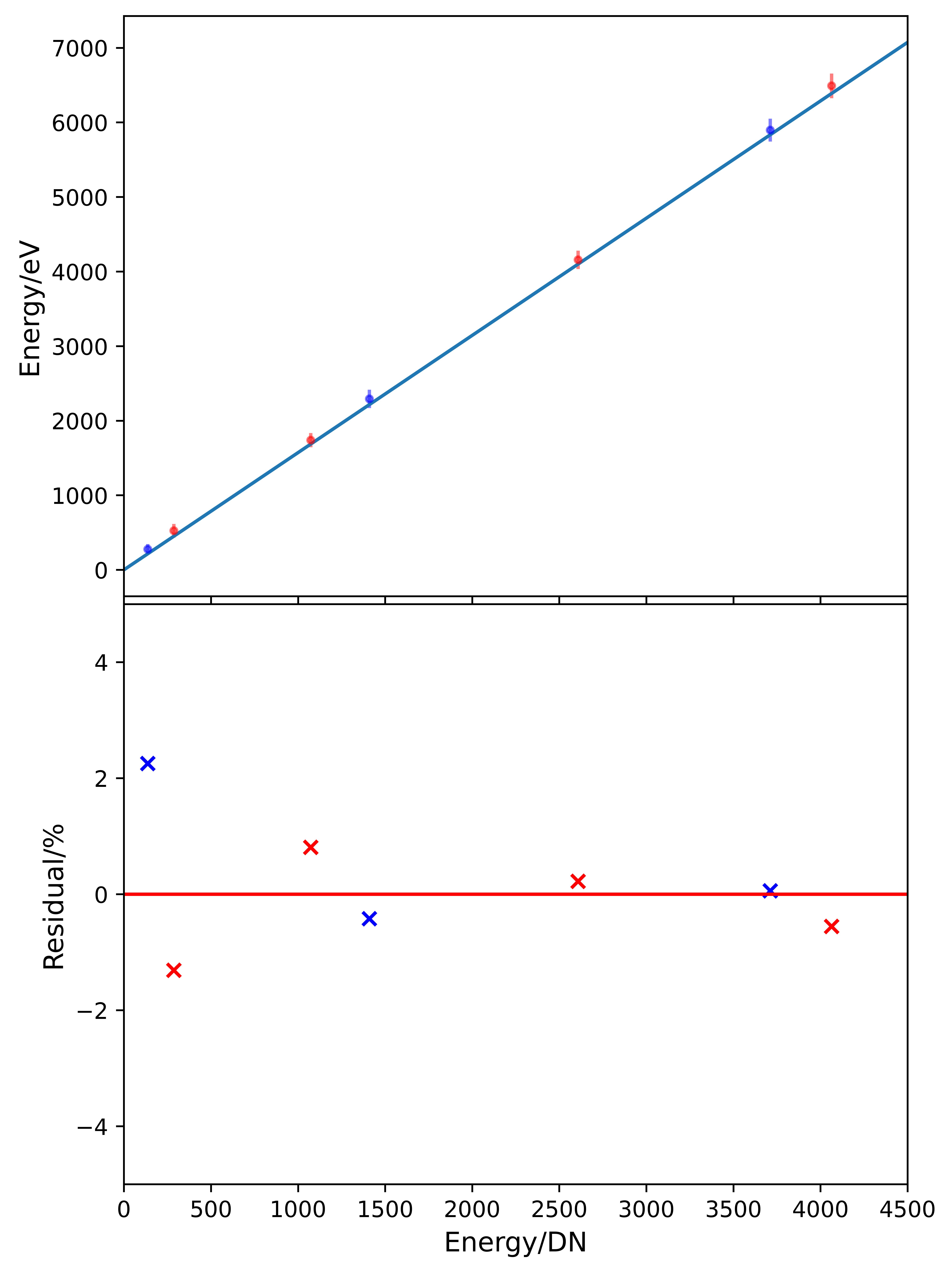}
        \caption{The energy calibration of G2020BSI. The blue dots indicate the main peak of sources including K$_{\alpha}$ line of Carbon, L$_{\alpha}$ line of Mo, and K$_{\alpha}$ line of Mn. The red dots indicate the other peak of sources including K$_{\alpha}$ line of O, L$_{\alpha}$ line of Si, K$_{\beta}$ line of Mn, and the Si escape peak of K$_{\alpha}$ line of Mn.}
    \label{Figure 6}
    \end{figure}
    As mentioned previously, it is necessary to discriminate between signals and noise for higher SNR. Generally, adjusting the threshold is an effective method. As shown in Figure 7, using the data processing flow above for images of the carbon target, the energy spectrum of the single-pixel events are plotted as yellow and blue lines at thresholds T$_{c}$ = 31 DN, and 62 DN, respectively. For single-pixel events of the carbon target, noise can be removed by setting a high threshold without sacrificing the number of signals. However, the most carbon-target events are multi-pixel events and for multi-pixel events, this method is limited. As shown in Figure 8, using the same data processing for images of Carbon target, energy spectrum of the all over-threshold events are plotted as the yellow, green and blue lines at the threshold T$_{c}$ = 31 DN, 45 DN, 62 DN, respectively. For multi-pixel events, discriminating the characteristic carbon peak from the noise peak is difficult when the threshold is low, and the number of signals is sacrificed when the threshold is high. Therefore, setting a reasonable threshold value is difficult. \par
    \begin{figure}[hbt]
        \centering
        \includegraphics[width=8cm]{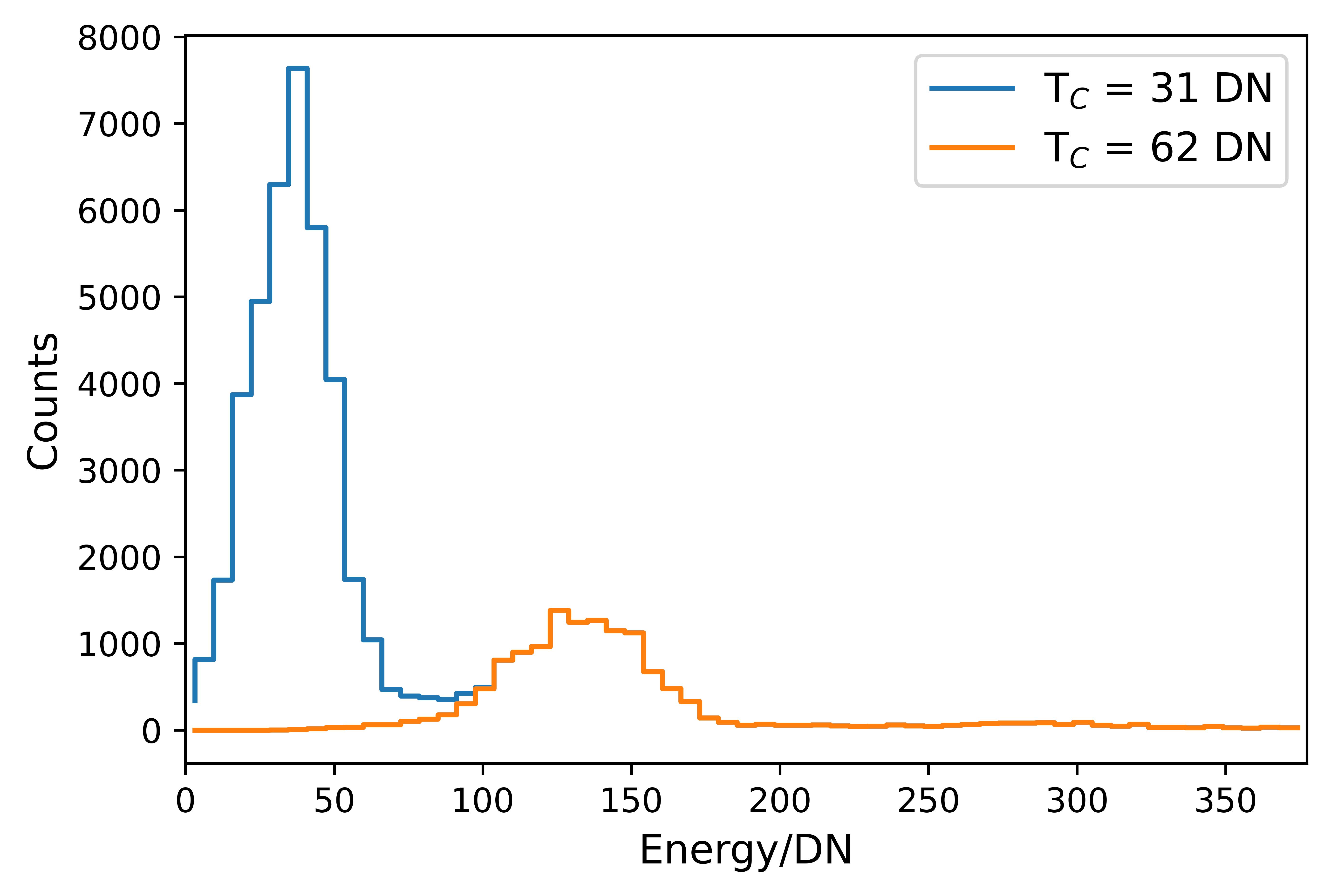}
        \caption{The energy spectrum of the carbon-target single-pixel events at the threshold T$_{c}$ = 31 DN and 62 DN.}\label{Figure 7}
    \end{figure}
    \begin{figure}[hbt]
        \centering
        \includegraphics[width=8cm]{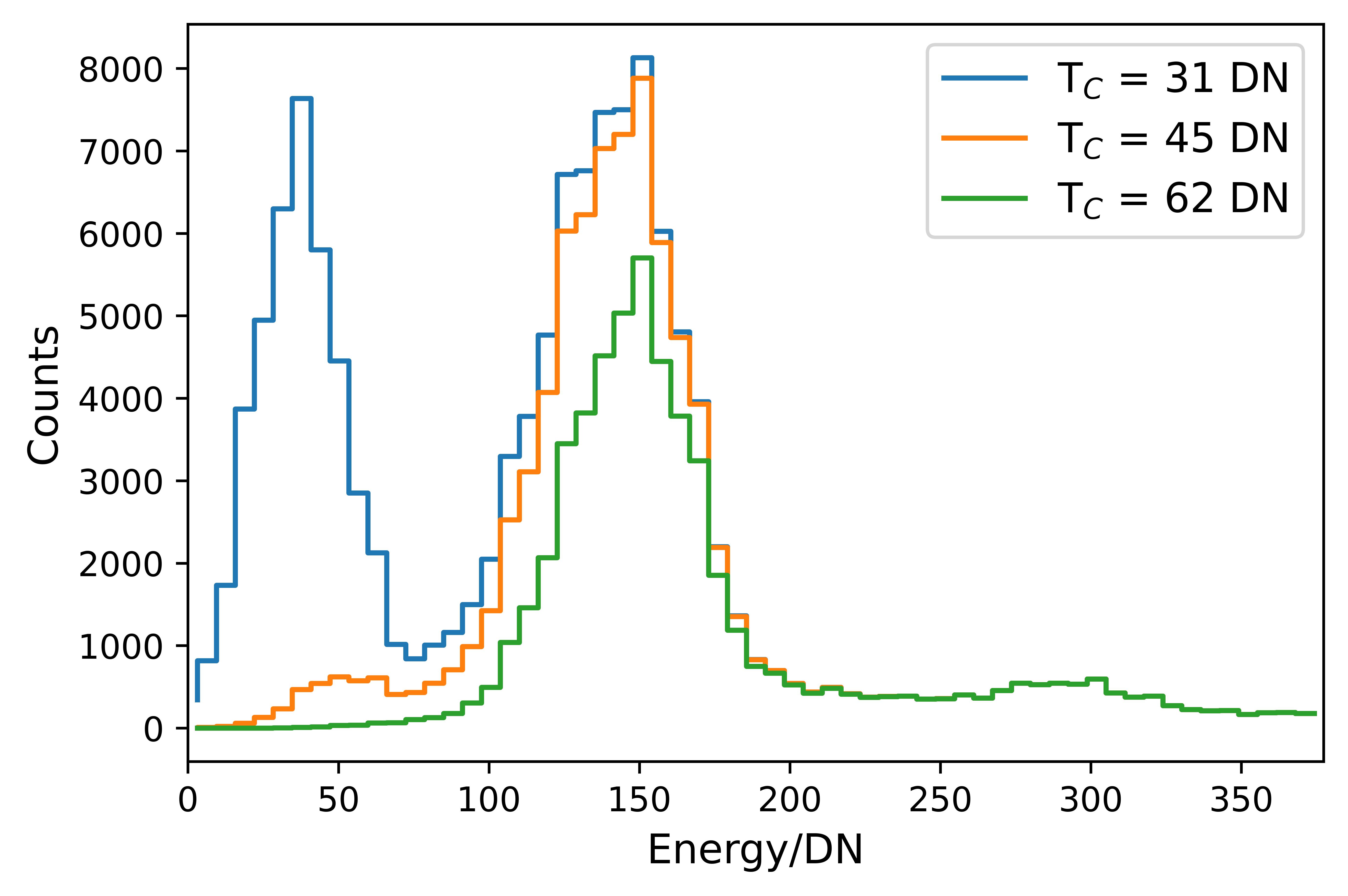}
        \caption{The energy spectrum of the all carbon-target over-threshold events at the threshold T$_{c}$ = 31 DN, 45 DN, and 62 DN.}\label{Figure 8}
    \end{figure}
    Therefore, a new discrimination method is needed for the discrimination of low energy soft X-ray events like K$_{\alpha}$ line of carbon without sacrificing the number of signals. One possible method to discriminate the signal is based on the charge distribution of G2020BSI. Theoretically, for signals, there is a correlation between the value of the center pixel (energy of the incident pixel) and the sum of the other eight pixel values (energy diffusing out of the incident pixel). For the false-triggered noise, there was no theoretical correlation between the value of the center pixel and the sum of the other eight pixel values. For convenience, two datasets were created. Using the data processing flow for images of the carbon target, all over-threshold events at threshold T$_{c}$ = 31 DN were used as the mixed dataset. Using the same data processing flow for images of noise, all over-threshold events at threshold T$_{c}$ = 31 DN were used as the noise dataset. The charge distribution could be reflected by a two-dimensional distribution with the center pixel value as the x-axis parameter and the sum of the other eight pixel values as the y-axis parameter. The two-dimensional distribution of the mixed dataset and noise dataset are shown as Figure 9a and 9b. It is worth mentioning that this charge distribution should be common for CMOS detectors based on the standard 4T pixel architecture and not only for GSENSEs.\par
    \begin{figure}[hbt]
        \centering
        \includegraphics[width=8cm]{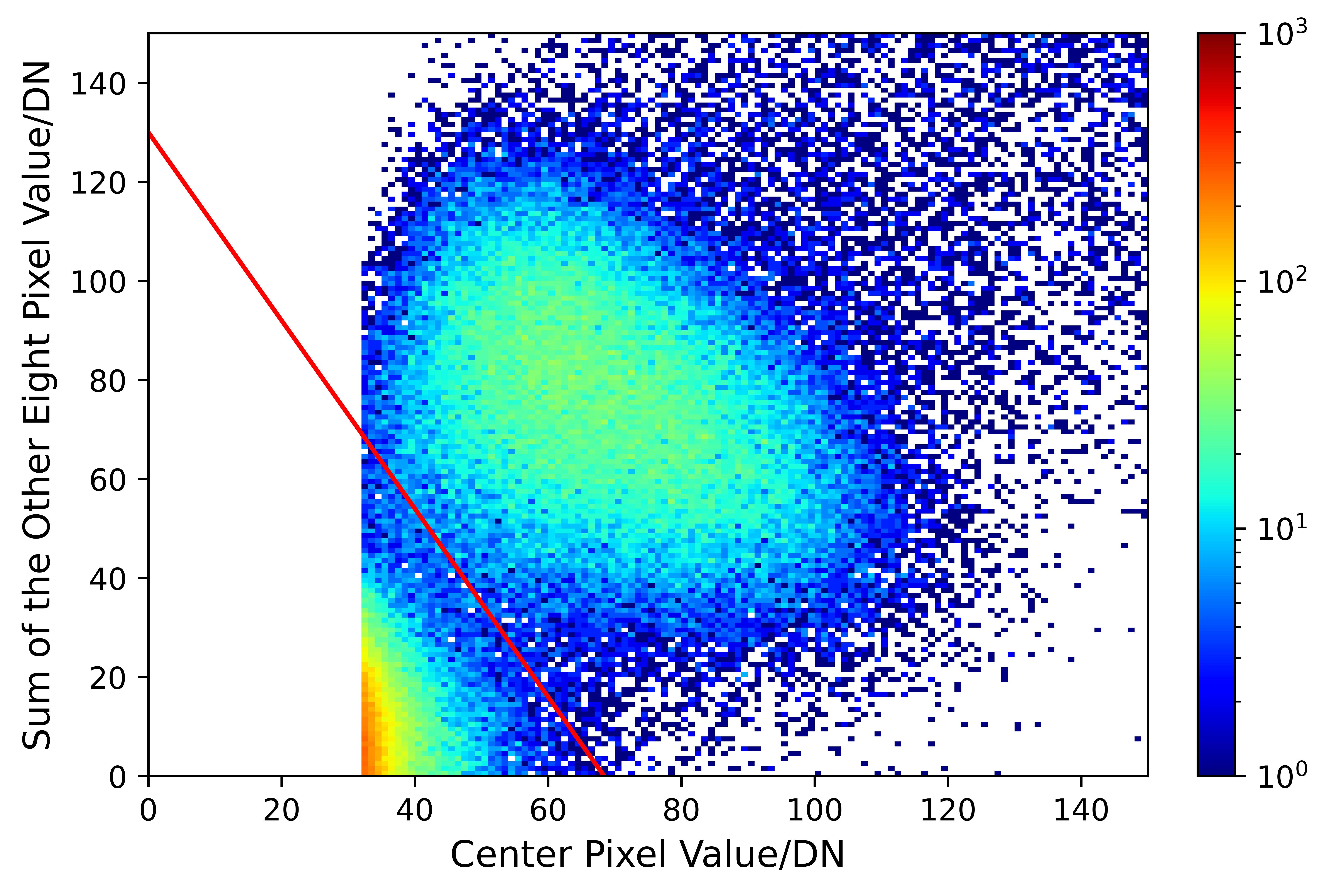}\label{(a}
        \centering
        \includegraphics[width=8cm]{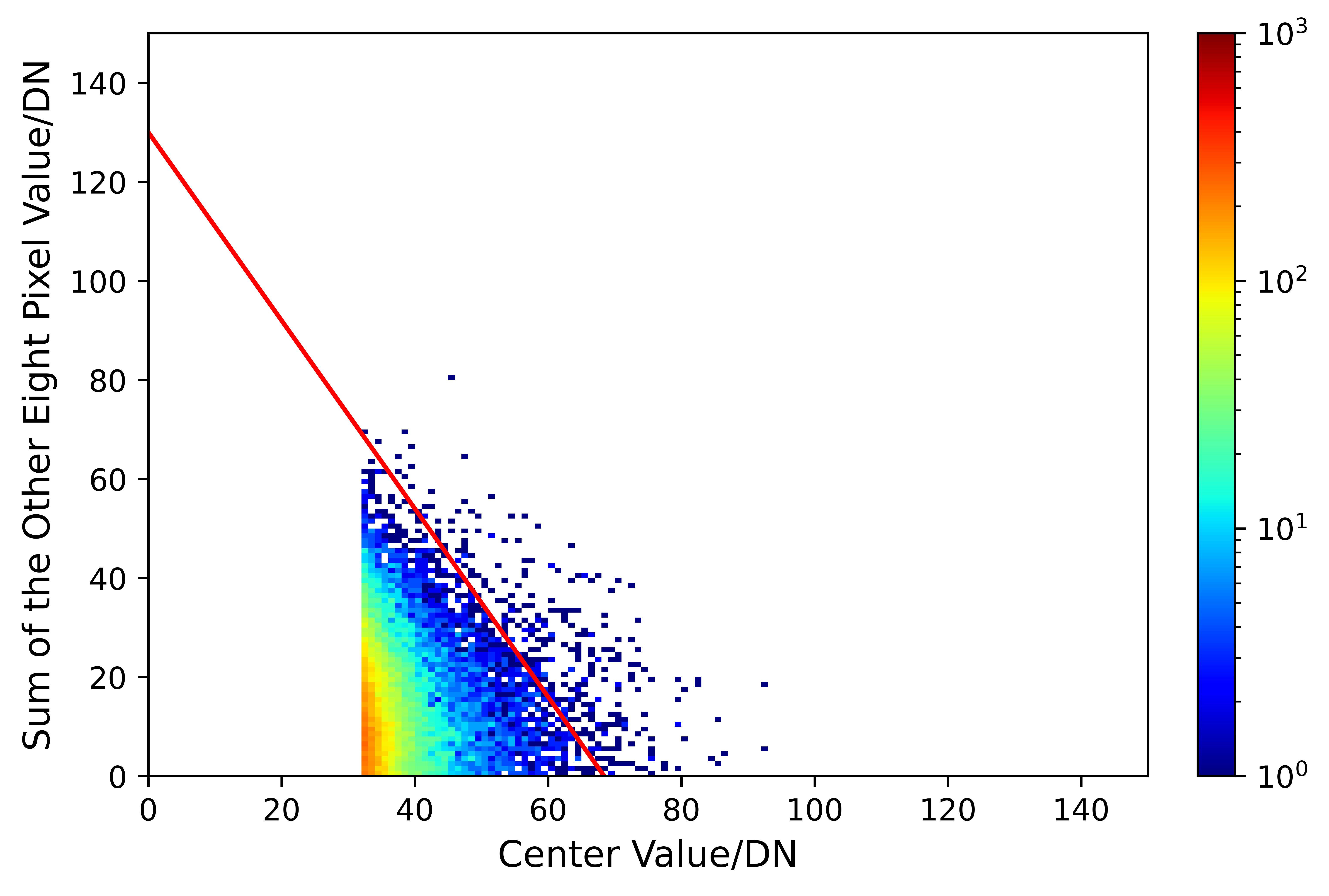}\label{(b}
        \caption{The two-dimensional distribution of the center pixel value vs. the sum of the other eight pixel values. a) the mixed dataset. b) the noise dataset.The red line represents the segmentation line $y\ =\ -1.9x\ +\ 130$. The events at the bottom left of the segmentation line are recognized as noise and the events at the top right of the segmentation line are conserdered as signals.}\label{Figure 9}
    \end{figure}

\section{Two-dimensional Segmentation Method and Optimization}
    Figure 9 showns that there is a difference between the charge distributions of the signal and noise. The red line $y\ =\ -1.9x\ +\ 130$ was initially selected from experience to segment the signal and noise in Figure 9, where the events at the top right of the red line were considered as signals and the events at the bottom left of the red line were considered as noise. This method which we called as two-dimensional segmentation method discriminates signals by segmenting a two-dimensional distribution with a straight line. We decided to use the figure of merit (FOM) \cite{Gamage201178} to evaluate the performance of the method. The next step is to calculate the FOM value.\par
    Using the same data processing flow for images of carbon target, all over-threshold events at threshold T$_{c}$ = 31 DN were used as the mixed dataset and the two-dimensional distribution of the mixed dataset was obtained. Suppose the segmentation line is $y\ =\ kx\ +\ b$. Calculate the distance $dis$ in the two-dimensional distribution to the segmentation line for each over-threshold event in the mixed dataset. The distribution of $dis$ clearly exhibited two peaks: a signal peak and a noise peak. The distribution was fitted using a double Gaussian function to obtain the peak position and full width at half maximum (2.355 × sigma $\sigma$). The FOM value could then be calculated using Equation (2) \cite{Gamage201178}. The larger the FOM value, the more the two peaks are separated, and the better the discrimination between the signal and noise. The maximum FOM value is obtained by scanning the slope of the line k. The value of b only affects the ratio of the signal (noise) before and after discrimination. In other words, the maximum FOM characterizes the performance of the method in discriminating signals.\par
    In this study, we only considered the discrimination of the signal and noise using a two-dimensional distribution with the center pixel values as the x-axis parameter. The y-axis parameter of the two-dimensional distribution is defined as the feature parameter of the two-dimensional segmentation method. The method discussed thus far is a two-dimensional segmentation with the sum of the other eight pixel values as the feature parameter. We verified that the sum of the other eight pixel values is a good feature parameter. However, methods that use other feature parameters may exhibit better discrimination performance. We considered some possible feature parameters. These were the second largest pixel value (SV), the second plus third largest pixel total values (SPTV), the standard deviation of the other eight pixel values (STDOV), the difference between the maximum and minimum values of the other eight pixel values (MMMOV), and the sum of the other eight pixel values (TOV). Using this process, the maximum FOM values for the methods using each feature parameter are listed in Table 2.
    \begin{equation}\label{Equation (2)}
        \centering
        FOM\ =\ \frac{|{Peak}_{1}-{Peak}_{2}|}{{FWHM}_{1}+{FWHM}_{2}}\ = \ \frac{|{Peak}_{1}-{Peak}_{2}|}{2.355({\sigma}_{1}+{\sigma}_{2})}
    \end{equation}
    \begin{table}[hbt]
        \centering
        \caption{The effect of the two-dimensional segmentation method with different feature parameters.}\label{Table 2}
        \begin{tabular}{cccc}
        \hline
        Feature Parameter & k     & b    & Max FOM   \\ \hline
        SV                & -0.56 & 40.9   & 1.086 \\
        SPTV              & -0.85 & 70.1   & 1.262 \\
        STDOV              & -0.22 & 14.9   & 1.027 \\
        MMMOV              & -0.87 & 57.0   & 0.917 \\
        TOV                & -2.18 & 124.5  & 1.032 \\ \hline
        \end{tabular}
    \end{table}
    According to Table 2, SPTV is the best feature parameter of two-dimensional segmentation method for G2020BSI among the five feature parameters. With SPTV as the feature parameter, the two-dimensional distribution of the center pixel value vs. SPTV is shown in Figure 10, and the distribution of the $dis$ from all over-threshold events to the segmentation line is shown in Figure 11.
    \begin{figure}[hbt]
        \centering
        \includegraphics[width=8cm]{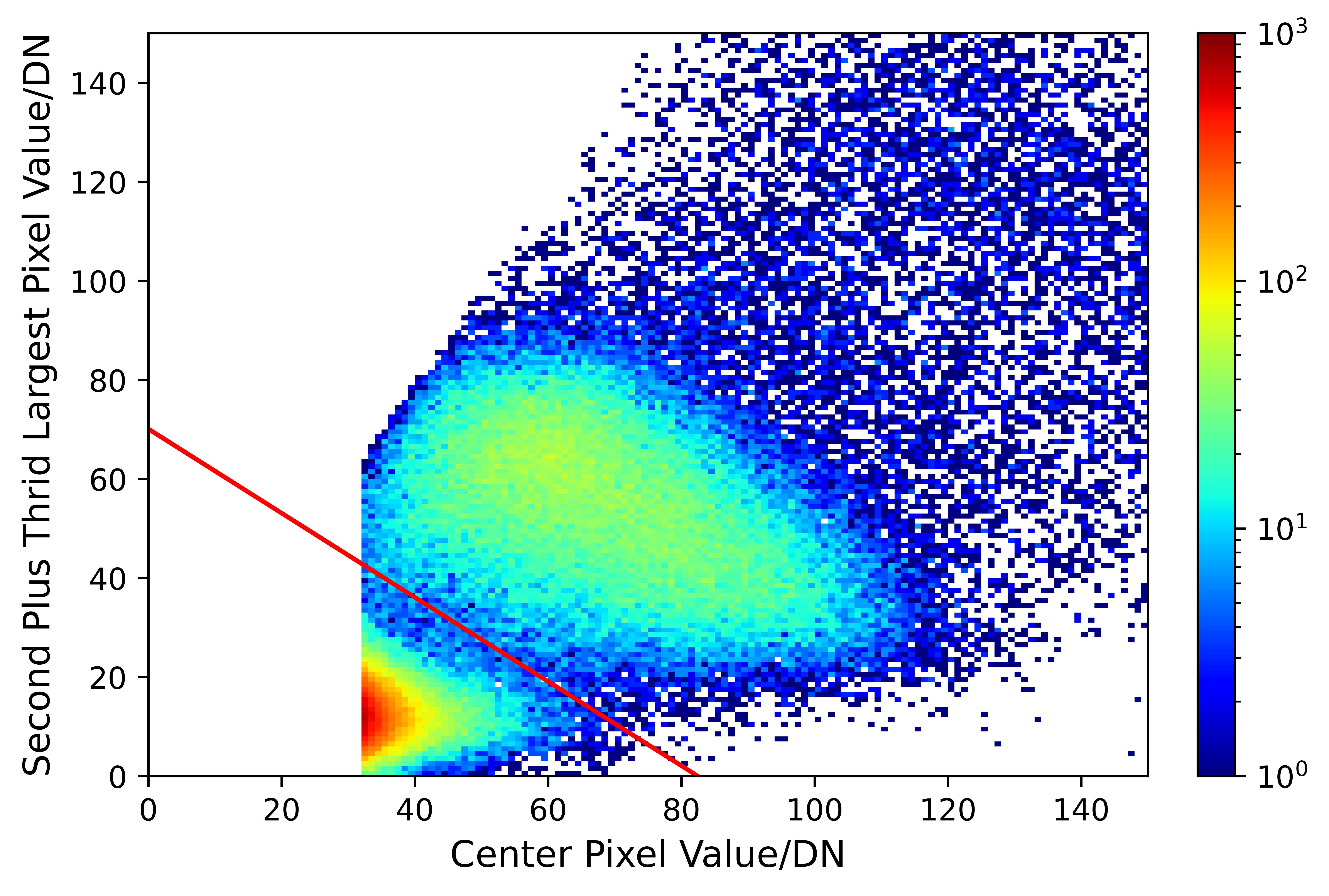}
        \caption{The two-dimensional distribution of the center pixel value vs. the second plus third largest pixel total value for events in the mixed data set.The red line is the segmentation line $y\ =\ -0.85x\ +\ 70.1$. The events at the bottom left of the segmentation line are recognized as noise and the events at the top right of the segmentation line are recognized as signals.}\label{Figure 10}
    \end{figure}
    \begin{figure}[hbt]
        \centering
        \includegraphics[width=8cm]{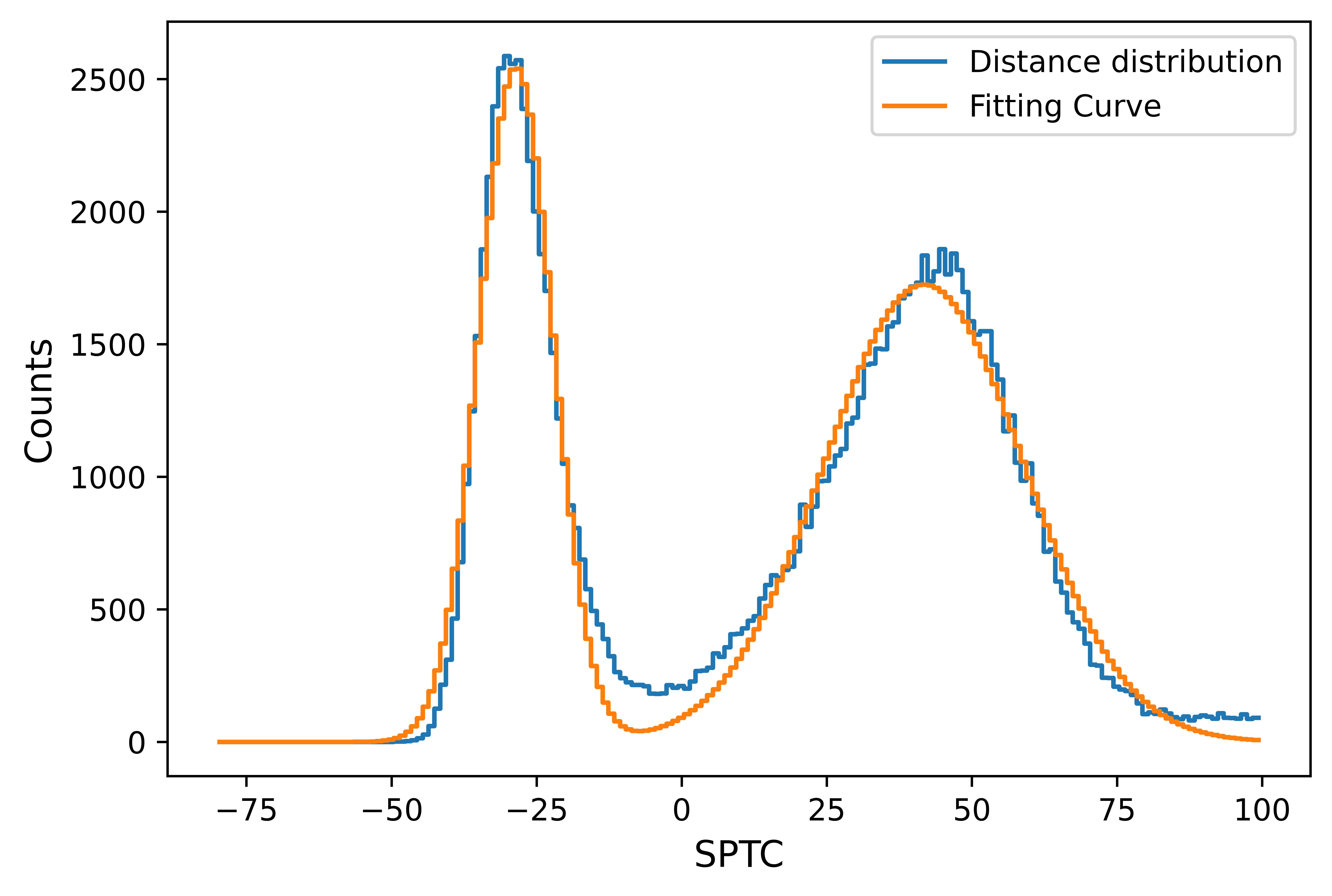}
        \caption{The distribution of the distances from all over-threshold events to the segmentation line. The distribution is fitted with a double Gaussian function for the calculation of FOM.}\label{Figure 11}
    \end{figure}
\section{Results and Discussion}
    The graphical results of the two-dimensional segmentation method when processing the low energy soft X-ray spectrum are shown as follow. For convenience, the energy spectrum of the mixed dataset is noted as the mixed energy spectrum. The energy spectrum of the noise dataset is denoted as the noise energy spectrum. The statistical signal energy spectrum was obtained by subtracting the signal energy spectrum from the mixed energy spectrum and is denoted as the subtraction energy spectrum. Here, the peak amplitude of the noise energy spectrum was normalized. The energy spectrum of the signal dataset obtained by processing the mixed dataset through the two-dimensional segmentation method is denoted as the two-dimensional segmentation energy spectrum. \par
    For the two-dimensional segmentation method with SPTV as the feature parameter and $y\ =\ -0.85x\ +\ 70.1$ as the segmentation line, the aforementioned three energy spectrum are shown in Figure 12. The two-dimensional segmentation energy spectrum was consistent with the subtraction energy spectrum within the error tolerance. Moreover, as presented in Figure 9 and Table 2, it can be expected that the two-dimensional segmentation method is not sensitive to parameters. That is, it is not necessary to determine the optimal parameters for the detector to obtain sufficiently good results. For the two-dimensional segmentation method with SV as the feature parameter and $y\ =\ -1.9x\ +\ 130$ as the segmentation line, the three aforementioned energy spectrum are shown in Figure 13. Evdiently, it still works sufficiently well at this point, even if the segmentation line at this point is clearly different from the scanned optimal segmentation line.\par
    \begin{figure}[hbt]
        \centering
        \includegraphics[width=8cm]{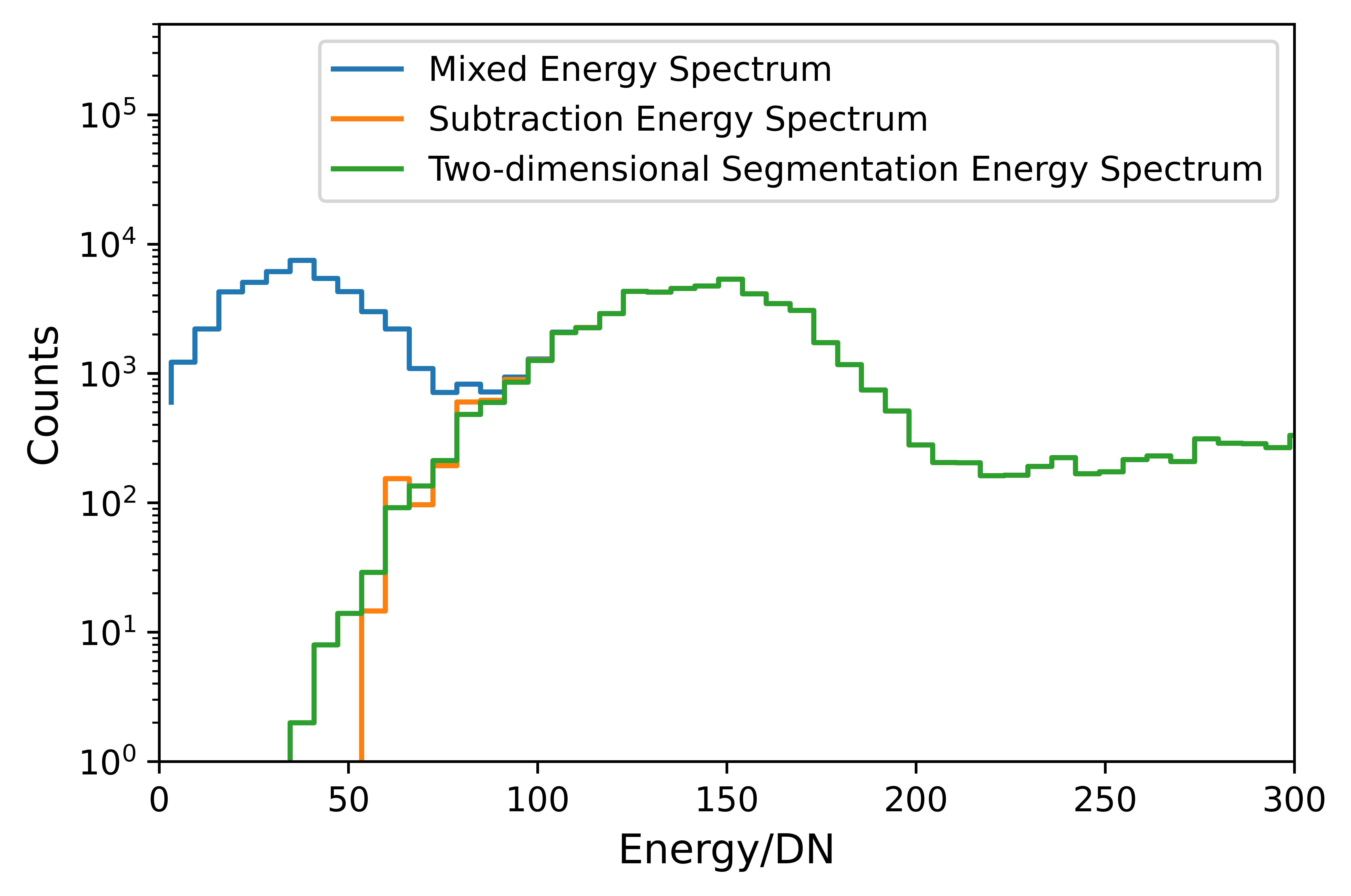}
        \caption{The mixedenergy spectrum with different processes. The spectrum shown by blue line is obtained without processing. The spectrum shown by orange line represents obtained by subtracting from the noise energy spectrum. The spectrum shown by green line is obtained by two-dimensional segmentation method with SPTV as the feature parameter and $y\ =\ -0.85x\ +\ 70.1$ as the segmentation line. }\label{Figure 12}
    \end{figure}
    \begin{figure}[hbt]
        \centering
        \includegraphics[width=8cm]{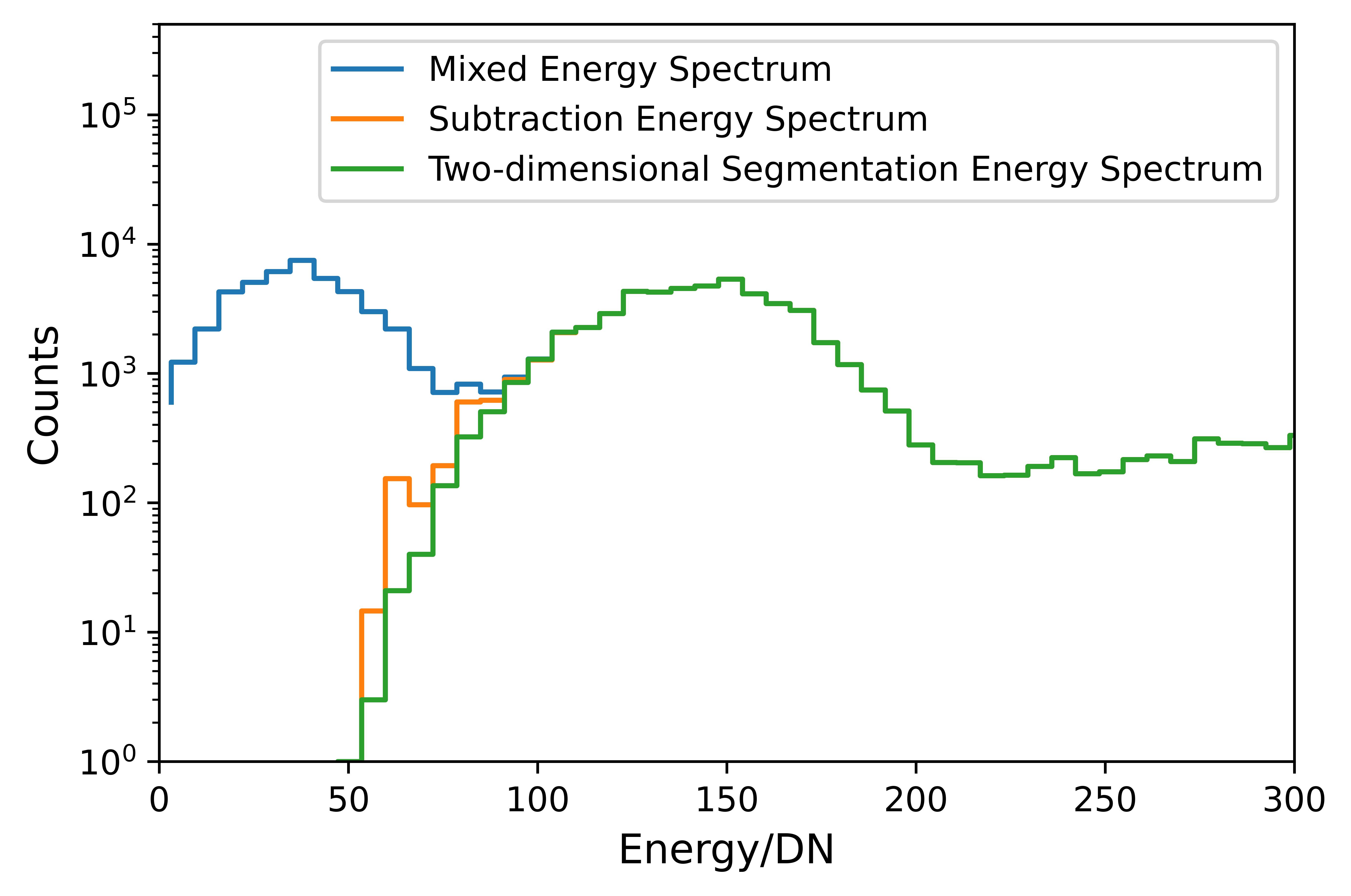}
        \caption{The mixedenergy spectrum with different processes. The spectrum shown by blue line is obtained without processing. The spectrum shown by orange line is obtained by subtracting from the noise energy spectrum. The spectrum shown by green line is obtained by two-dimensional segmentation method with TOV as the feature parameter and $y\ =\ -1.9x\ +\ 130$ as the segmentation line.}\label{Figure 13}
    \end{figure}
\section{Conclusion}
    In this study, a new noise discrimination method for CMOS detectors is proposed for low-energy X-ray events. The experimental results indicated that the two-dimensional segmentation method effectively reduces the lower energy threshold of the G2020BSI detector, almost without signal loss, and improves the SNR of low-energy X-ray events. In addition, this method is insensitive to feature parameters, and good results can be achieved when using empirically selected feature parameters.\par
    The two-dimensional segmentation method reduces the energy lower-threshold of the CMOS detector almost without signal loss. The G2020BSI detector can be used as a focal plane detector for the important soft X-ray bioimaging technology soft X-ray tomography owing to the energy lower-threshold lower than energy of carbon K$_{\alpha}$ line\cite{Mille2022}, for soft X-ray astronomy to improve the MDP\cite{Weisskopf2010} etc. In addition, considering that the charge distribution model should be common for the standard 4T pixel architecture, the two-dimensional segmentation method could be applied to other CMOS detectors based on the 4T pixel architecture, and even other pixel-type detectors. In the future, we intend to apply the two-dimensional segmentation method to other pixel-type detectors to verify its applicability.
        








\bibliography{ref}

\begin{thebibliography}{18}
\expandafter\ifx\csname natexlab\endcsname\relax\def\natexlab#1{#1}\fi
\providecommand{\url}[1]{\texttt{#1}}
\providecommand{\href}[2]{#2}
\providecommand{\path}[1]{#1}
\providecommand{\DOIprefix}{doi:}
\providecommand{\ArXivprefix}{arXiv:}
\providecommand{\URLprefix}{URL: }
\providecommand{\Pubmedprefix}{pmid:}
\providecommand{\doi}[1]{\href{http://dx.doi.org/#1}{\path{#1}}}
\providecommand{\Pubmed}[1]{\href{pmid:#1}{\path{#1}}}
\providecommand{\bibinfo}[2]{#2}
\ifx\xfnm\relax \def\xfnm[#1]{\unskip,\space#1}\fi
\bibitem[{Harada et~al.(2019{\natexlab{a}})Harada, Teranishi, Watanabe, Zhou,
  Yang, Bogaerts, and Wang}]{Harada_2019}
\bibinfo{author}{T.~Harada}, \bibinfo{author}{N.~Teranishi},
  \bibinfo{author}{T.~Watanabe}, \bibinfo{author}{Q.~Zhou},
  \bibinfo{author}{X.~Yang}, \bibinfo{author}{J.~Bogaerts},
  \bibinfo{author}{X.~Wang},
\newblock \bibinfo{title}{Energy- and spatial-resolved detection using a
  backside-illuminated cmos sensor in the soft x-ray region},
\newblock \bibinfo{journal}{Applied Physics Express} \bibinfo{volume}{12}
  (\bibinfo{year}{2019}{\natexlab{a}}) \bibinfo{pages}{082012}.
  \DOIprefix\doi{10.7567/1882-0786/ab32d2}.
\bibitem[{Harada et~al.(2019{\natexlab{b}})Harada, Teranishi, Watanabe, Zhou,
  Bogaerts, and Wang}]{Harada_2020}
\bibinfo{author}{T.~Harada}, \bibinfo{author}{N.~Teranishi},
  \bibinfo{author}{T.~Watanabe}, \bibinfo{author}{Q.~Zhou},
  \bibinfo{author}{J.~Bogaerts}, \bibinfo{author}{X.~Wang},
\newblock \bibinfo{title}{High-exposure-durability, high-quantum-efficiency
  ($>$ 90\%) backside-illuminated soft-x-ray cmos sensor},
\newblock \bibinfo{journal}{Applied Physics Express} \bibinfo{volume}{13}
  (\bibinfo{year}{2019}{\natexlab{b}}) \bibinfo{pages}{016502}.
  \DOIprefix\doi{10.7567/1882-0786/ab5b5e}.
\bibitem[{Cooper et~al.(2020)Cooper, Wise, Woodward, and Varagnat}]{Cooper2020}
\bibinfo{author}{J.~T. Cooper}, \bibinfo{author}{A.~J. Wise},
  \bibinfo{author}{T.~Woodward}, \bibinfo{author}{A.~Varagnat},
\newblock \bibinfo{title}{{Custom high-sensitivity CCD and sCMOS detectors for
  high-harmonic generation, x-ray absorption spectroscopy, soft x-ray
  microscopy/tomography, and hard x-ray detection}}  (\bibinfo{year}{2020})
  \bibinfo{pages}{56}. \DOIprefix\doi{10.1117/12.2546096}.
\bibitem[{Desjardins et~al.(2020)Desjardins, Medjoubi, Sacchi, Popescu,
  Gaudemer, Belkhou, Stanescu, Swaraj, Besson, Vijayakumar, Pautard,
  Noureddine, Mercere, {Da Silva}, Orsini, Menneglier, and
  Jaouen}]{Desjardins2020}
\bibinfo{author}{K.~Desjardins}, \bibinfo{author}{K.~Medjoubi},
  \bibinfo{author}{M.~Sacchi}, \bibinfo{author}{H.~Popescu},
  \bibinfo{author}{R.~Gaudemer}, \bibinfo{author}{R.~Belkhou},
  \bibinfo{author}{S.~Stanescu}, \bibinfo{author}{S.~Swaraj},
  \bibinfo{author}{A.~Besson}, \bibinfo{author}{J.~Vijayakumar},
  \bibinfo{author}{S.~Pautard}, \bibinfo{author}{A.~Noureddine},
  \bibinfo{author}{P.~Mercere}, \bibinfo{author}{P.~{Da Silva}},
  \bibinfo{author}{F.~Orsini}, \bibinfo{author}{C.~Menneglier},
  \bibinfo{author}{N.~Jaouen},
\newblock \bibinfo{title}{{Backside-illuminated scientific CMOS detector for
  soft X-ray resonant scattering and ptychography}},
\newblock \bibinfo{journal}{Journal of Synchrotron Radiation}
  \bibinfo{volume}{27} (\bibinfo{year}{2020}) \bibinfo{pages}{1577--1589}.
  \DOIprefix\doi{10.1107/S160057752001262X}.
\bibitem[{nosuke Ishikawa et~al.(2018)nosuke Ishikawa, Takahashi, Watanabe,
  Narukage, Miyazaki, Orita, Takeda, Nomachi, Fujishiro, and
  Hodoshima}]{Ishikawa2018191}
\bibinfo{author}{S.~nosuke Ishikawa}, \bibinfo{author}{T.~Takahashi},
  \bibinfo{author}{S.~Watanabe}, \bibinfo{author}{N.~Narukage},
  \bibinfo{author}{S.~Miyazaki}, \bibinfo{author}{T.~Orita},
  \bibinfo{author}{S.~Takeda}, \bibinfo{author}{M.~Nomachi},
  \bibinfo{author}{I.~Fujishiro}, \bibinfo{author}{F.~Hodoshima},
\newblock \bibinfo{title}{High-speed x-ray imaging spectroscopy system with
  zynq soc for solar observations},
\newblock \bibinfo{journal}{Nuclear Instruments and Methods in Physics Research
  Section A: Accelerators, Spectrometers, Detectors and Associated Equipment}
  \bibinfo{volume}{912} (\bibinfo{year}{2018}) \bibinfo{pages}{191--194}.
  \DOIprefix\doi{10.1016/j.nima.2017.11.033}.
\bibitem[{Heymes et~al.(2020)Heymes, Stefanov, Soman, Gorret, Hall, Minoglou,
  Morris, Pratlong, Prod'homme, Tsiolis, and Holland}]{Julian2020}
\bibinfo{author}{J.~Heymes}, \bibinfo{author}{K.~Stefanov},
  \bibinfo{author}{M.~Soman}, \bibinfo{author}{D.~Gorret},
  \bibinfo{author}{D.~Hall}, \bibinfo{author}{K.~Minoglou},
  \bibinfo{author}{D.~Morris}, \bibinfo{author}{J.~Pratlong},
  \bibinfo{author}{T.~Prod'homme}, \bibinfo{author}{G.~Tsiolis},
  \bibinfo{author}{A.~Holland},
\newblock \bibinfo{title}{{Development of a photon-counting near-fano-limited
  x-ray CMOS image sensor for THESEUS' SXI}},
\newblock in: \bibinfo{editor}{A.~D. Holland}, \bibinfo{editor}{J.~Beletic}
  (Eds.), \bibinfo{booktitle}{X-Ray, Optical, and Infrared Detectors for
  Astronomy IX}, volume \bibinfo{volume}{11454},
  \bibinfo{organization}{International Society for Optics and Photonics},
  \bibinfo{publisher}{SPIE}, \bibinfo{year}{2020}, p. \bibinfo{pages}{114540I}.
  \DOIprefix\doi{10.1117/12.2560162}.
\bibitem[{Narukage et~al.(2020)Narukage, nosuke Ishikawa, Sakao, and
  Wang}]{Narukage2020}
\bibinfo{author}{N.~Narukage}, \bibinfo{author}{S.~nosuke Ishikawa},
  \bibinfo{author}{T.~Sakao}, \bibinfo{author}{X.~Wang},
\newblock \bibinfo{title}{{High-speed back-illuminated CMOS sensor for
  photon-counting-type imaging-spectroscopy in the soft X-ray range}},
\newblock \bibinfo{journal}{Nuclear Instruments and Methods in Physics
  Research, Section A: Accelerators, Spectrometers, Detectors and Associated
  Equipment} \bibinfo{volume}{950} (\bibinfo{year}{2020})
  \bibinfo{pages}{162974}. \DOIprefix\doi{10.1016/j.nima.2019.162974}.
\bibitem[{Wu et~al.(2023{\natexlab{a}})Wu, Ling, Wang, Zhang, Yuan, and
  Zhang}]{Wu2023a}
\bibinfo{author}{Q.~Wu}, \bibinfo{author}{Z.~Ling}, \bibinfo{author}{X.~Wang},
  \bibinfo{author}{C.~Zhang}, \bibinfo{author}{W.~Yuan}, \bibinfo{author}{S.~N.
  Zhang},
\newblock \bibinfo{title}{{Improving the X-Ray Energy Resolution of a
  Scientific CMOS Detector by Pixel-level Gain Correction}},
\newblock \bibinfo{journal}{Publications of the Astronomical Society of the
  Pacific} \bibinfo{volume}{135} (\bibinfo{year}{2023}{\natexlab{a}}).
  \DOIprefix\doi{10.1088/1538-3873/acbcf3}.
  \href{http://arxiv.org/abs/2303.01027}{{\tt arXiv:2303.01027}}.
\bibitem[{Wu et~al.(2023{\natexlab{b}})Wu, Ling, Zhang, Zhou, Wang, Yuan, and
  Zhang}]{Wu2023}
\bibinfo{author}{Q.~Wu}, \bibinfo{author}{Z.~Ling}, \bibinfo{author}{C.~Zhang},
  \bibinfo{author}{Q.~Zhou}, \bibinfo{author}{X.~Wang},
  \bibinfo{author}{W.~Yuan}, \bibinfo{author}{S.~N. Zhang},
\newblock \bibinfo{title}{{Investigating the image lag of a scientific CMOS
  sensor in X-ray detection}},
\newblock \bibinfo{journal}{Nuclear Instruments and Methods in Physics
  Research, Section A: Accelerators, Spectrometers, Detectors and Associated
  Equipment} \bibinfo{volume}{1050} (\bibinfo{year}{2023}{\natexlab{b}})
  \bibinfo{pages}{168180}. \DOIprefix\doi{10.1016/j.nima.2023.168180}.
  \href{http://arxiv.org/abs/2303.08425}{{\tt arXiv:2303.08425}}.
\bibitem[{Weisskopf et~al.(2010)Weisskopf, Elsner, and O'Dell}]{Weisskopf2010}
\bibinfo{author}{M.~C. Weisskopf}, \bibinfo{author}{R.~F. Elsner},
  \bibinfo{author}{S.~L. O'Dell},
\newblock \bibinfo{title}{{On understanding the figures of merit for detection
  and measurement of x-ray polarization}},
\newblock \bibinfo{journal}{Space Telescopes and Instrumentation 2010:
  Ultraviolet to Gamma Ray} \bibinfo{volume}{7732} (\bibinfo{year}{2010})
  \bibinfo{pages}{77320E}. \DOIprefix\doi{10.1117/12.857357}.
\bibitem[{Ogino et~al.(2021)Ogino, Arimoto, Sawano, Yonetoku, Yu, Watanabe,
  Hiraga, Yuhi, Hatori, Kume, and Hasegawa}]{Ogino2021}
\bibinfo{author}{N.~Ogino}, \bibinfo{author}{M.~Arimoto},
  \bibinfo{author}{T.~Sawano}, \bibinfo{author}{D.~Yonetoku},
  \bibinfo{author}{P.~Yu}, \bibinfo{author}{S.~Watanabe},
  \bibinfo{author}{J.~S. Hiraga}, \bibinfo{author}{D.~Yuhi},
  \bibinfo{author}{S.~Hatori}, \bibinfo{author}{K.~Kume},
  \bibinfo{author}{T.~Hasegawa},
\newblock \bibinfo{title}{{Performance verification of next-generation Si CMOS
  soft X-ray detector for space applications}},
\newblock \bibinfo{journal}{Nuclear Instruments and Methods in Physics
  Research, Section A: Accelerators, Spectrometers, Detectors and Associated
  Equipment} \bibinfo{volume}{987} (\bibinfo{year}{2021})
  \bibinfo{pages}{164843}. \DOIprefix\doi{10.1016/j.nima.2020.164843}.
\bibitem[{Chen et~al.(2022)Chen, Wang, Xu, Zhao, Qiu, Hou, Yang, Ma, Chen,
  Zhao, Liu, Zhao, and Zhu}]{Chen2022}
\bibinfo{author}{C.~Chen}, \bibinfo{author}{Y.~Wang}, \bibinfo{author}{Y.~Xu},
  \bibinfo{author}{Z.~Zhao}, \bibinfo{author}{H.~Qiu},
  \bibinfo{author}{D.~Hou}, \bibinfo{author}{X.~Yang}, \bibinfo{author}{J.~Ma},
  \bibinfo{author}{Y.~Chen}, \bibinfo{author}{Y.~Zhao},
  \bibinfo{author}{H.~Liu}, \bibinfo{author}{X.~Zhao},
  \bibinfo{author}{Y.~Zhu},
\newblock \bibinfo{title}{{Performance of a focal plane detector for soft X-ray
  imaging spectroscopy based on back-illuminated sCMOS}},
\newblock \bibinfo{journal}{Nuclear Instruments and Methods in Physics
  Research, Section A: Accelerators, Spectrometers, Detectors and Associated
  Equipment} \bibinfo{volume}{1030} (\bibinfo{year}{2022})
  \bibinfo{pages}{166465}. \DOIprefix\doi{10.1016/j.nima.2022.166465}.
  \href{http://arxiv.org/abs/2111.11610}{{\tt arXiv:2111.11610}}.
\bibitem[{Mille et~al.(2022)Mille, Yuan, Vijayakumar, Stanescu, Swaraj,
  Desjardins, Favre-Nicolin, Belkhou, and Hitchcock}]{Mille2022}
\bibinfo{author}{N.~Mille}, \bibinfo{author}{H.~Yuan},
  \bibinfo{author}{J.~Vijayakumar}, \bibinfo{author}{S.~Stanescu},
  \bibinfo{author}{S.~Swaraj}, \bibinfo{author}{K.~Desjardins},
  \bibinfo{author}{V.~Favre-Nicolin}, \bibinfo{author}{R.~Belkhou},
  \bibinfo{author}{A.~P. Hitchcock},
\newblock \bibinfo{title}{Ptychography at the carbon k-edge},
\newblock \bibinfo{journal}{Communications Materials} \bibinfo{volume}{3}
  (\bibinfo{year}{2022}) \bibinfo{pages}{8}.
  \DOIprefix\doi{10.1038/s43246-022-00232-8}.
\bibitem[{Zhang et~al.(2022)Zhang, Hu, Xin, Li, Guo, Xue, Dong, Yu, Wang, Lei,
  and Geng}]{Zhang2022}
\bibinfo{author}{B.~Zhang}, \bibinfo{author}{C.~Hu}, \bibinfo{author}{Y.~Xin},
  \bibinfo{author}{Y.~Li}, \bibinfo{author}{Z.~Guo}, \bibinfo{author}{Z.~Xue},
  \bibinfo{author}{L.~Dong}, \bibinfo{author}{S.~Yu},
  \bibinfo{author}{X.~Wang}, \bibinfo{author}{S.~Lei},
  \bibinfo{author}{L.~Geng},
\newblock \bibinfo{title}{{MOS-based model of four-transistor CMOS image sensor
  pixels for photoelectric simulation}},
\newblock \bibinfo{journal}{Chinese Physics B} \bibinfo{volume}{31}
  (\bibinfo{year}{2022}). \DOIprefix\doi{10.1088/1674-1056/ac3819}.
\bibitem[{Heymes et~al.(2022)Heymes, Ivory, Stefanov, Buggey, Hetherington,
  Soman, and Holland}]{Heymes2022}
\bibinfo{author}{J.~Heymes}, \bibinfo{author}{J.~Ivory},
  \bibinfo{author}{K.~Stefanov}, \bibinfo{author}{T.~Buggey},
  \bibinfo{author}{O.~Hetherington}, \bibinfo{author}{M.~Soman},
  \bibinfo{author}{A.~Holland},
\newblock \bibinfo{title}{{Characterisation of a soft X-ray optimised CMOS
  Image Sensor}},
\newblock \bibinfo{journal}{Journal of Instrumentation} \bibinfo{volume}{17}
  (\bibinfo{year}{2022}). \DOIprefix\doi{10.1088/1748-0221/17/05/P05003}.
\bibitem[{Ma et~al.(2015)Ma, Liu, Li, Zhou, Chang, and Wang}]{Ma2015}
\bibinfo{author}{C.~Ma}, \bibinfo{author}{Y.~Liu}, \bibinfo{author}{J.~Li},
  \bibinfo{author}{Q.~Zhou}, \bibinfo{author}{Y.~Chang},
  \bibinfo{author}{X.~Wang},
\newblock \bibinfo{title}{{A 4MP high-dynamic-range, low-noise CMOS image
  sensor}},
\newblock \bibinfo{journal}{Image Sensors and Imaging Systems 2015}
  \bibinfo{volume}{9403} (\bibinfo{year}{2015}) \bibinfo{pages}{940305}.
  \DOIprefix\doi{10.1117/12.2083085}.
\bibitem[{Wang et~al.(2019)Wang, Ling, Zhang, Jia, Wang, Wu, Yuan, and
  Zhang}]{Wang2019}
\bibinfo{author}{W.~X. Wang}, \bibinfo{author}{Z.~X. Ling},
  \bibinfo{author}{C.~Zhang}, \bibinfo{author}{Z.~Q. Jia},
  \bibinfo{author}{X.~Y. Wang}, \bibinfo{author}{Q.~Wu}, \bibinfo{author}{W.~M.
  Yuan}, \bibinfo{author}{S.~N. Zhang},
\newblock \bibinfo{title}{{Characterization of a BSI sCMOS for soft X-ray
  imaging spectroscopy}},
\newblock \bibinfo{journal}{Journal of Instrumentation} \bibinfo{volume}{14}
  (\bibinfo{year}{2019}). \DOIprefix\doi{10.1088/1748-0221/14/02/P02025}.
\bibitem[{Gamage et~al.(2011)Gamage, Joyce, and Hawkes}]{Gamage201178}
\bibinfo{author}{K.~Gamage}, \bibinfo{author}{M.~Joyce},
  \bibinfo{author}{N.~Hawkes},
\newblock \bibinfo{title}{A comparison of four different digital algorithms for
  pulse-shape discrimination in fast scintillators},
\newblock \bibinfo{journal}{Nuclear Instruments and Methods in Physics
  Research, Section A: Accelerators, Spectrometers, Detectors and Associated
  Equipment} \bibinfo{volume}{642} (\bibinfo{year}{2011}) \bibinfo{pages}{78
  – 83}. \DOIprefix\doi{10.1016/j.nima.2011.03.065}.

\end{thebibliography}
\bibliographystyle{elsarticle-num-names}

\end{document}